\documentstyle[12pt,aaspp4]{article}

\def\Mo {$M_{\odot}$}

\def\MJ {$M_{J}$}

\def\ms {m s$^{-1}$}

\def\ups{$\upsilon$ And}
\def\hd{HD~10697}
\def\degr{^{\circ}}
\def\Msini{$M \sin i$}

\begin{document}

\title{ Analysis of the Hipparcos Measurements of \hd ---\\
A Mass Determination of a Brown-Dwarf Secondary}

\author{Shay Zucker and Tsevi Mazeh}
\affil{School of Physics and Astronomy, Raymond and Beverly Sackler
Faculty of Exact Sciences, Tel Aviv University, Tel Aviv, Israel\\
shay, mazeh@wise.tau.ac.il}

\begin{abstract}

\hd\ is a nearby main-sequence star around which a planet candidate has
 recently been discovered by means of radial-velocity measurements (Vogt
 et al.~1999, submitted to ApJ). The stellar orbit has a period of about
 three years, the secondary minimum mass is 6.35 Jupiter masses (\MJ)
 and the minimum semi-major axis is 0.36 milli-arc-sec (mas).

Using the Hipparcos data of \hd\ {\it together} with the spectroscopic
elements of Vogt et al.~(1999) we found a semi-major axis of $2.1\pm 0.7$
mas, implying a mass of $38\pm 13$ \MJ\ for the unseen companion. We
therefore suggest that the secondary of \hd\ is probably a brown dwarf,
orbiting around its parent star at a distance of 2 AU.

\subjectheadings{astrometry --- planetary systems --- stars: individual
(\hd)}
\end{abstract}

\section{INTRODUCTION}

More than twenty candidates for extrasolar planets have been announced
over the past four years (e.g., Mayor \& Queloz 1995; Noyes et al.\
1997; Marcy \& Butler 1998; Marcy, Cochran \& Mayor 2000).  In each
case, very precise stellar radial-velocity measurements, with a
precision of 10~m~s$^{-1}$ or better, indicated the presence of a
low-mass unseen companion orbiting a nearby solar-type star. These
high-precision discoveries came almost a decade after the first 'planet
candidate' around HD 114762 was discovered (Latham et al.\ 1989; Mazeh,
Latham \& Stefanik 1996) with much lower precision. In most cases, the
individual masses of the companions are not known, because the
inclination angles of their orbital planes relative to our line of sight
cannot be derived from the spectroscopic data. The minimum masses for
all candidates, attained for an inclination angle of $90^{\circ}$, are
in the range 0.5 to 10 Jupiter masses (\MJ).

To derive the actual mass we need additional information, like precise
astrometry of the orbit, from which we can derive the inclination, and
therefore the secondary mass. This can be done at least for the cases
where the primary mass can be estimated from its spectral type. At
present, the astronomical community has at hand the accurate astrometric
Hipparcos data, which have already yielded numerous orbits with small
semi-major axes (ESA 1997; S\"oderhjelm 1999), down to a few
milli-arc-sec (mas). Perryman et al.~(1996) indeed used the Hipparcos
data to constrain the orbits of some planet candidates.

The Hipparcos data, which span about three years, can be most effective
for systems with periods comparable to the lifetime of the satellite
mission. As the time base of the precise radial-velocity surveys is getting
longer, we expect more such planets to be discovered. Such an example was
the outermost planet of \ups\ whose period was found to be close to 3
years.  Mazeh et al.~(1999) used the Hipparcos data {\it together} with
the spectroscopic data of \ups\ to get a mass of $10.1^{+4.7}_{-4.6}$
\MJ\ for the outermost known planet of that system. That work has shown
the great potential of the combination of spectroscopic data and
Hipparcos measurements to derive masses for planet candidates.

Very recently Vogt et al.~(1999) detected radial-velocity evidence of
new six planet candidates. One of these stars is \hd, with a period of
$1072\pm 10$ days and a companion lower mass estimate of 6.35 \MJ. The
minimum semi-major axis of the stellar orbit was found to be 0.36
mas. In this Letter we follow Mazeh et al.~(1999) and combine the
spectroscopic elements with the Hipparcos data to find that the unseen
companion has a mass of about 40 \MJ. We therefore suggest that the secondary
is probably a brown dwarf in close orbit, of about 2 AU, around its
parent star.

\section{The Hipparcos Astrometry}  

\hd\ (109 Psc, HIP8159; $\alpha$=01:44:55.82,
$\delta$=+20:04:59.34 [J2000]; V = 6.3), is a G5IV star, whose mass is
estimated by Vogt et al.~(1999) to be 1.10 \Mo.  The Hipparcos data of
\hd\ included 33 independent abscissa measurements, which are the
results of work of FAST and NDAC --- the two consortia involved in
reduction of the raw Hipparcos data. These 33 measurements represent
data from 17 orbits of the satellite, analyzed separately by the two
consortia and then de-correlated to produce 2 independent measurements
out of each orbit (van Leeuwen 1997; van Leeuwen \& Evans 1998; Perryman
et al.~1997). In orbit no. 614 only the NDAC consortium produced a
measurement of \hd.

The present analysis used the spectroscopic elements of \hd\ as given by
Vogt et al.~(1999).  These included the spectroscopic period, $P=
1072\pm 10$ day, the periastron passage, $T_0= 2451482\pm 39$, the
radial-velocity amplitude, $K= 119\pm 3$ \ms, the eccentricity $e=
0.12\pm 0.02 $, and the longitude of the periastron, $\omega= 113\pm
14$. The orbital astrometric elements include $P, \, T_0, \, e, \,
\omega$ and three additional elements --- the semi-major axis, $a_1$,
the inclination, $i$, and the longitude of the nodes, $\Omega$. In
addition, the astrometric solution includes the five regular astrometric
parameters --- the parallax, the position (in right ascension and
declination) and the proper motion (in right ascension and
declination). All together we had 12 parameter model to fit to the
astrometric data, four of which are common with the spectroscopic orbit.

To find the best astrometric orbit we used the values of $P, \, T_0, \, e,
\, \omega$ as given by the spectroscopic orbit, and solved for the other
parameters. To do that we considered a dense grid on the ($a_1, i$) plane,
and found the values of the five regular astrometric parameters and
$\Omega$ that minimized the $\chi^2$ statistics for each value of ($a_1,
i$).  The result of this search is a $\chi^2$ {\it function}, which
depends on $a_1$ and $i$. The square-root normalized $\chi^2$ is plotted
in Figure~1 as a two-dimensional function.

The figure shows a very pronounced ``valley'' at about $a_1 = 2$ mas and
$i = 160\degr$, indicating a detection of an astrometric motion.  To
derive the best semi-major axis and inclination for the system we used
another spectroscopic element --- the radial-velocity amplitude $K$,
which has not been used so far in the analysis. This element induces a
constraint on the product of $a_1$ and $\sin i$, which for the \hd\ case
results in

\begin{equation}
a_1 \sin i= 0.36\pm0.01\
            \Bigl({{P}\over{1072\,{\rm day}}}\Bigr)
            \Bigl({{K}\over{119\,{\rm m\,s^{-1}}}}\Bigr)
            \Bigl({{\sqrt{1-e^2}}\over{0.993}}\Bigr)
            \Bigl({{\pi}\over{30.71\,{\rm mas}}}\Bigr)
            \ {\rm mas} \ ,
\end{equation}
where we have used here the Hipparcos {\it catalog} parallax,
$\pi=30.71\pm0.81$ mas. This constraint is plotted in Figure 1 as a
continuous line both on the ($a_1, i$) plane and on the square-root
normalized $\chi^2$ surface. The product was calculated assuming the
best-fit value of $\pi$ --- 30.3 mas. Using the catalog value yielded
very similar results. In Figure 2 we collapsed the
two-dimensional function onto the line of
Eq. (1). We can see a clear minimum at $170\degr$. This corresponds to a
semi-major axis of 2.1 mas and a mass of 38 \MJ. 
Detailed $\chi^2$ analysis resulted in

\begin{equation}
a_1= 2.1 \pm 0.7\,{\rm mas} \ \ ,\,\ M_{sec}=38\pm 13\,M_J \ ,
\end{equation}
where $M_{sec}$ is the mass of the unseen companion. 

\section{Discussion}

All the 29 planet candidates discovered so far have \Msini\ in the range
of 0.5--10 \MJ. No report has been published yet about any precise
radial-velocity detection of a secondary with minimum masses in the
range of 10--80 \MJ. The large radial-velocity survey of K and G stars
with lower precision of Mayor et al.~(1997) also yielded only few
binaries with minimum secondary mass in the range of 10--80 \MJ, most of
which are actually main-sequence secondaries with low inclination
(Halbwachs et al.~1999; Udry et al.~1999).  These studies attest to the
paucity of secondaries in the range of 10--80 \MJ, and suggest a
distinction between the population of stellar secondaries and planets
(Vogt et el. 1999; Mazeh 1999). This distinction might also help to find
the mass upper limit for planetary companions.

An upper limit of the planetary mass at about 10--20 \MJ\ is consistent
with the accumulated distribution of planetary masses (Mazeh 1999). It
is also consistent with the mass of the outermost known companion of
\ups\ --- $10 \pm 5$ \MJ\ (Mazeh et al.~1999).  It therefore seems that
the secondary of \hd\ mass is probably too large for a planetary
mass. We therefore suggest that the secondary of \hd\ is a brown dwarf
orbiting around its parent star at a distance of 2 AU.

The paradigm behind the distinction between brown dwarfs and planets
assumes that binaries, even with small-mass secondaries, are formed by a
different mechanism than that of planets (e.g., Boss 1996). Brown-dwarf
secondaries are therefore at the low-mass end of the distribution of
secondaries that were formed as binaries and not as planets. The study
of the lower end of the mass distribution has great importance to our
understanding of binary formation, specially if we compare them to the
emerging population of brown-dwarf field stars (e.g., B\`ejar, Zapatero
Osorio \& Rebolo 1999; Binney 1999; Tinney 1999 and see references
therein). Until now the paucity of brown dwarf secondaries in
short-period binaries could give the impression that such objects do not
exist. This work suggests that there are such secondaries, even though
rare, in this range of masses.

As usual, objects in binaries with short-enough periods supply the
precious opportunity of dynamical mass determination. The combination of
astrometry and radial-velocity work yields here a first mass
determination of a brown dwarf which is model independent. The
separation between the two components of \hd\ is about 70 mas (Vogt et
al.~1999). Ground-based interferometry might be able to resolve the two
objects despite the huge brightness difference between them and supply
for the first time brightness measurements of a brown dwarf with
dynamical mass determination.  In addition, time dependent spectroscopy
might help to obtain a spectrum of the faint companion, enabling us to
confront the theory of brown-dwarf evolutionary models (e.g., Burrows \&
Sharp 1999) with the observations.

This work was supported by the US-Israel Binational Science
Foundation through grant 97-00460 and the Israeli Science Foundation.

\section*{REFERENCES}
\begin{description}

\item B\`ejar, V.~J.~S., Zapatero Osorio, M.~R., \& Rebolo, R. 1999,
  ApJ, 521, 671

\item Binney, J. 1999, MNRAS, 307, L27

\item Boss, A.~P. 1996, Nature, 379, 397 

\item Burrows, A., \& Sharp, C.~M. 1999, ApJ, 512, 843

\item Halbwachs, J.~L., Arenou, F., Mayor, M., Udry, S., \& Queloz,
  D. 1999 A\&A, submitted

\item Latham, D.~W., Mazeh, T., Stefanik, R.~P., Mayor, M., \& Burki,
G. 1989, Nature, 339, 38

\item ESA 1997, The Hipparcos and Tycho Catalogues, ESA SP-1200

\item van Leeuwen, F. 1997, Space Sc. Rev., 81, 201

\item van Leeuwen, F., \& Evans, D.~W. 1998, A\&AS, 130, 157

\item Marcy, G.~W., \& Butler R.~P. 1998, ARAA, 36, 57

\item Marcy, G.~W., Cochran, W.~D., \& Mayor, M. 2000, in Protostars and
Planets IV, ed.  V. Mannings, A. P. Boss \& S. S. Russell (Tucson:
University of Arizona Press), in press
 
\item Mayor, M., \& Queloz, D. 1995, Nature, 378, 355 

\item Mayor, M., Queloz, D., Udry, S., \& Halbwachs, J.-L. 1997, in IAU
Coll. 161, Astronomical and Biochemical Origins and Search for Life in
the Universe ed. C. B. Cosmovici, S. Boyer, \& D. Werthimer 
(Bolognia: Editrice Compositori) 313

\item Mazeh, T. 1999, in IAU Coll. 170, Precise Stellar Radial
  Velocities, eds. J.B. Hearnshaw \& C.D. Scarfe, in press

\item Mazeh, T., Latham, D.~W., \& Stefanik R.~P. 1996, ApJ, 466, 415 

\item Mazeh, T., Zucker, S., Dalla Torre, A., \& van Leeuwen, F., 1999,
  ApJL, 522, L149

\item Noyes, R.~W., Jha, S., Korzennik, S.~G., Krockenberger, M.,
Nisenson, P., Brown, T.~M., Kennelly, E.~J., \& Horner, S.~D. 1997,
ApJL, 483, L111

\item Perryman, M.~A.~C., et al.~1996, A\&A, 310, L21

\item Perryman, M.~A.~C., et al.~1997, A\&A, 323, L49

\item S\"oderhjelm, S. 1999, A\&A, 341, 121

\item Tinney, C.~G. 1999, Nature, 397, 37

\item Vogt, S.~S., Marcy, G.~W., Butler, R.~P., Apps, K. 1999, ApJ, submitted

\item Udry, S., Mayor, M., Naef, D., Pepe, F., Queloz, D., Santos, N.,
  Burnet, M. 1999, A\&A, submitted
\end{description}

\begin{center}
FIGURE LEGENDS
\end{center}

\figcaption{The minimum square-root normalized $\chi^2$ statistics 
as a function of $ a_1$ and $i$. 
The continuous line is the $a_1 \sin i= 0.36\ {\rm mas}$ constraint.}

\figcaption{The minimum square-root normalized $\chi^2$ statistics as a
function of $i$, given the constraint $a_1 \sin i= 0.36\ {\rm
mas}$.}

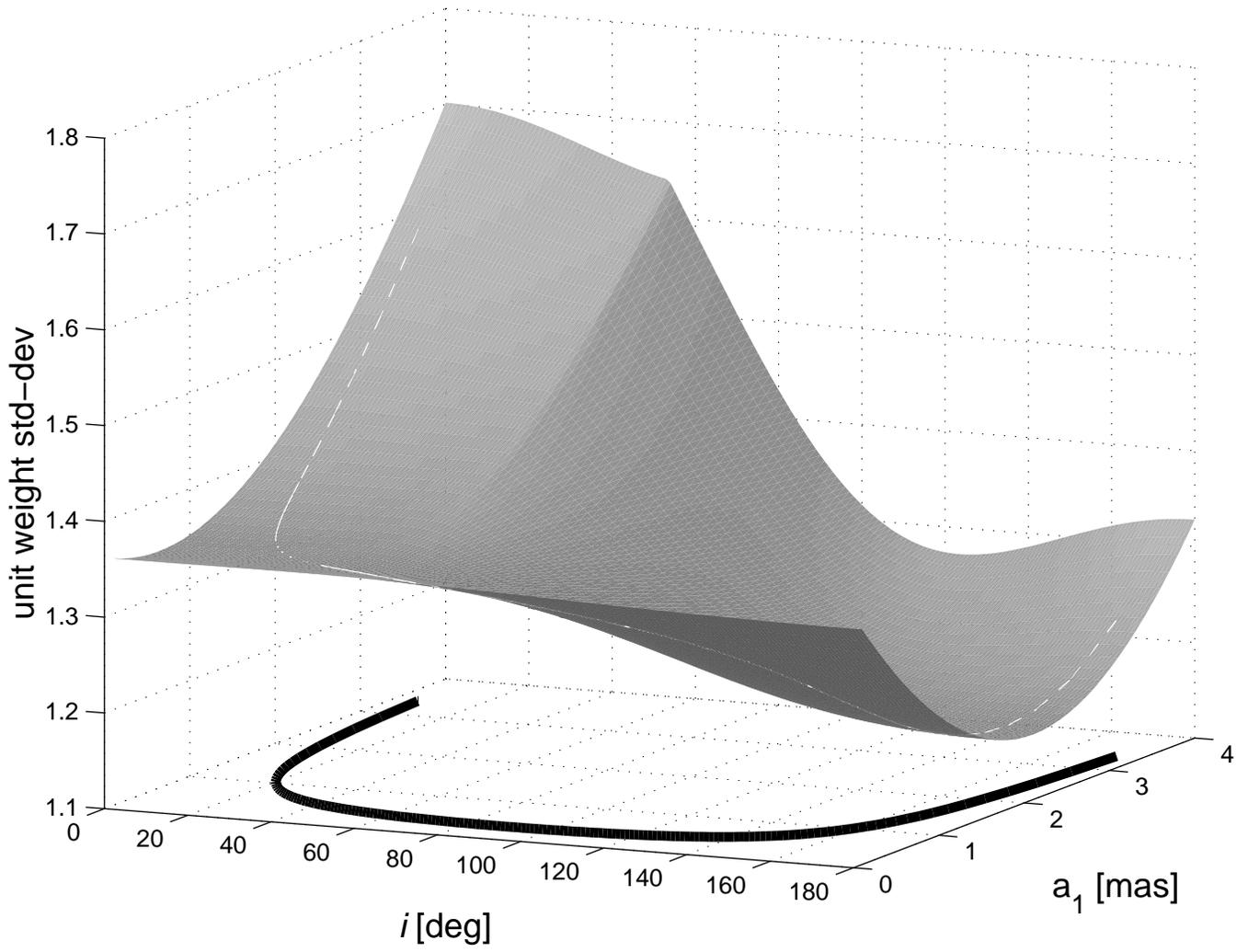
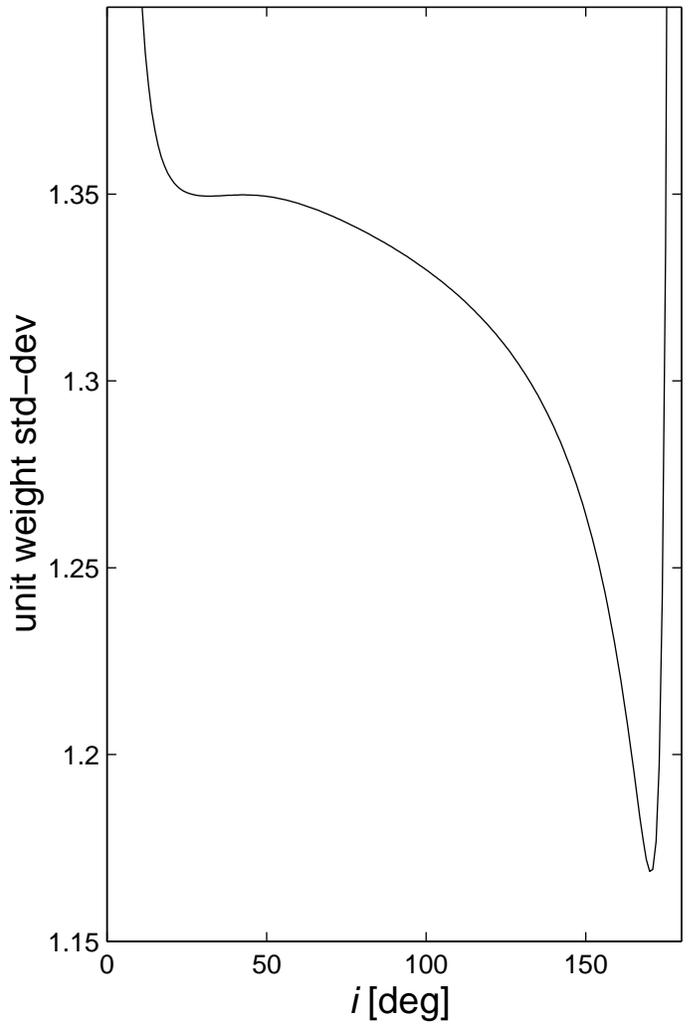
\end{document}